# Decoding the Gender Gap: Addressing Gender Stereotypes and Psychological Barriers to Empower Women in Technology

**Zahra Fakoor Harehdasht**
**Independent Researcher, Düsseldorf, Germany**
za.fakoor@gmail.com

**Raziyeh Saki**
**Independent Researcher, Tehran, Iran**
raziyeh.saki158@gmail.com

## Abstract

Recently, the unequal presence of women compared to men in technology has attracted the attention of researchers and practitioners across multiple fields. It is time to regard this problem as a global crisis that not only limits access to talent but also reduces the diversity of perspectives that shape technological innovation. This article examines the psychological and social barriers that influence this gap, as well as the interventions designed to reduce it. Using a structured review, the findings assemble evidence on the role of early gender stereotypes in the family and school and the continuation of this crisis in educational and career choices, through to the psychological challenges women face in professional settings, such as feelings of self-undervaluation, occupational anxiety, a heightened fear of technology, and structural limitations in educational environments. Special attention is paid to Germany, where the technology gap is particularly evident and where multiple national programs have been implemented to address it. The present review shows that effective solutions require more than anti-discrimination policies: they should include educational practices, organizational reforms, mentoring, and psychological support. The article concludes by outlining practical and research implications and introduces the NEURON project as a pilot interdisciplinary initiative aimed at accelerating current empowerment efforts and developing new programs for women in technology occupations.

# Introduction

In the past few decades, technology has become a primary driver of economic growth and new products. Jobs in software, data, and related fields have spread fast and now compete for skilled workers. Still, women remain underrepresented across many parts of the tech economy. This gender gap has significant implications, it is not only a missed chance to use more talent, but it also reduces the variety of perspectives that shape the design and purpose of technological products (Trinkenreich et al., 2022).

Global numbers show only slow progress. Women's participation in science and engineering has risen in some places, but in fields like artificial intelligence and data science their share is often far lower than in other fields (Lazzaroni & Pal, 2024). Some commentators have even called the scale of the deficit a crisis; the World Economic Forum's 2024 estimate that gender parity in the workplace will not be reached for more than 130 years at current rates, this makes plain how entrenched the problem can be (World Economic Forum, 2024).

The reason that this gap persists lays on barriers such as hiring discrimination or wage inequality, and hidden layers of challenges also play a crucial role in shaping women's professional trajectories. But there are also forces that discourage women's participation long before they reach the hiring stage. Many girls grow up with the mindset that skills in science, technology, engineering, and mathematics (STEM) fields are "male domains," a belief reinforced in families and schools. They may hear adults saying that "technology is hard" in ways that specifically discourage girls. For them, it is rare to encounter women in leadership roles within education or technology-related professions. Over time, such stereotypes cause girls to be less encouraged to pursue computer science or engineering, which amplifies feelings of self-doubt and mistrust in their own abilities (Miller et al., 2024; Wang & Degol, 2017).

Entering the workplace can bring another set of pressures. A common experience reported by women in tech is the so-called impostor feeling, which means even when their work is solid, they doubt their competence and attribute success to luck. Repeated social experiences such as being interrupted in meetings, having credit taken for ideas or being steered away from technical tasks reinforce a lack of confidence. This is referred to as "technophobia," a dual fear that arises not only from the complexity of technology but also from the culture and stereotypes (Trinkenreich et al., 2022).

Fixing the gap therefore calls for more than eliminating discrimination or changing hiring rules. Practical steps include clearer career paths after parental leave, visible senior women as mentors, and classroom practices that give all students hands-on experience with coding and hardware. Psychological supports matter too: mentoring, peer networks, and feedback practices that help people recognize real accomplishments reduce chronic self-doubt.

This paper brings together published studies on the psychological and social barriers women face in technology and the intervention strategies designed to address them. We use a structured literature review to map what the evidence says, where the gap is, and finally which types of interventions have measurable effects. The goal is to create an evidence-based foundation to inform practical interventions in Germany, where the gender gap in technology is particularly noticeable (Lazzaroni & Pal, 2024). This research aims to translate global insights into actionable strategies, to provide a roadmap for initiatives seeking to close the gender gap and ensure that women's presence becomes an integral part of the technology industry's future in Germany.

To achieve this goal, our structured literature review will answer the following primary research questions on a global scale:

- **Research Question 1 (RQ1): How do gender stereotypes hinder women from entering and persisting in technology careers?**

- **Research Question 2 (RQ2): How do psychological barriers impact women's success and persistence in technology?**

- **Research Question 3 (RQ3): Which empowerment strategies are most effective at enabling women to overcome stereotypes and barriers to increase their participation and success in technology?**

The evidence synthesized from these three research questions will be used to answer a core guiding question which is for the application of this research in Germany:

- **How can the evidence-based insights from this structured literature review be translated into actionable strategies to address the specific context of the gender gap in the German technology sector?**

# Theoretical Framework

This chapter presents a working framework for understanding why a gender gap persists in technology: how broad social forces feed into personal experience, and how those experiences shape what interventions can realistically accomplish. We start by locating gender stereotypes as a social foundation, then we trace how those norms become psychological barriers at the level of feeling and beliefs, then finally, we address empowerment as a layered process to provides theoretical grounding for assessing the effectiveness of interventions and solutions.

**Descriptive and prescriptive gender stereotypes**

Stereotypes lie at the root of the gender gap in technology in the pervasive influence of gender stereotypes. Gender stereotypes are defined academically as generalized beliefs about the attributes of the women and men. According to Madeline E. Heilman (2012), these stereotypes operate through two distinct mechanisms, that are referred to as descriptive stereotypes and prescriptive stereotypes. Descriptive stereotypes determine how women and men are believed to be like, this creates a perceived lack of fit for women in technical fields, since it associates the technical skills with masculinity. The second mechanism, prescriptive stereotypes act as social norms, it dictates how women should behave. Women who violate these norms for instance by being ambitious, they often face being pushed back at work. This results in a situation where they face negative consequences and instead of being competent they are judged negatively for it. In the study, gender stereotypes are emphasized as the most prominent factor underlying the gender gap. These stereotypes emerge from a complex interplay of social, cultural, and educational factors and they cause persistent psychological effects on women.

**Psychological barriers: belonging and self-efficacy**

The gender stereotypes in technology are not just social beliefs; they create psychological barriers that hinder women from entering and persisting in this field. These psychological barriers are the internal and cognitive or emotional consequences of an environment with stereotypes. These barriers manifest mainly through two interconnected mechanisms. First, a sense of not belonging arises when the environment gives the signals that women do not fit withing the masculine culture of tech. Second, this lack of fit signal weakens women's confidence and undermines their belief in their own abilities, even when their skills are just as strong as men. Feeling like they don't belong and doubting their abilities are two key ways that social stereotypes push women out of technology fields or hinder them from entering this field (Master et al., 2016). In this study, two dimensions of psychological barriers are identified, the psychological vulnerability of women which results from being excluded from the field and the psychological barriers experienced by women who are already working in this domain. Together, these factors prevent women's participation and success in this domain.

**Defining empowerment: a multi-dimensional process**

To overcome the impact of stereotypes and psychological barriers and empower women, this study looks for empowering methods and solution to address these challenges. Based on Zimmerman (1995), empowerment is defined as a multi-dimensional process that through it individuals gain control over their lives, which involves three components. This includes an interpersonal component, where a woman builds self-efficacy and overcomes internalized barriers, an interactional component, where she develops the awareness and skills to navigate the system, and a behavioral component, where she participates actively in shaping her career. This framework provides the theoretical basis to evaluate methods like mentorship and role modeling that are effective, because they target all three components of the empowerment process. Throughout this research, after identifying the main barriers, the focus is on two key areas: raising awareness of how gender stereotypes and psychological barriers affect women and creating practical empowerment strategies that can help reduce the gender gap in technology.

**Guiding Theories**

To explore how stereotypes shape career interest, self-efficacy, and belonging, this paper draws on Expectancy-Value Theory (EVT). Originating with Jacquelynne Eccles, EVT argues that people decide to engage with and persist in activities based on two things: how likely they think they are to succeed, and how much the activity matters to them subjectively (Wang & Degol, 2013). In technology fields, gender stereotypes negatively impact both components of this model. When tech is framed as a "masculine domain," for example, women may reasonably downgrade their expectations for success; when the stereotype that tech is "geeky" and socially isolating make the work seem less meaningful, which reduces interest and belonging. EVT provides the main theoretical framework for Research Question 1, which aims to measure how societal stereotypes affect women's motivation to pursue a career in technology.

Stereotype Threat Theory supplies the mechanism explaining how psychological barriers translate into poorer performance, lower persistence, and reduce aspirations. Steele and Aronson described the experience in which people worry they might confirm a negative stereotype about their group, a pressure that places a significant cognitive load on an individual and raise anxiety during demanding tasks (Steele & Aronson, 1995). The immediate effect is often measurable in momentary performance drops; the longer-term consequence (after repeated exposures) can be a decision to avoid or leave that domain. That said, some scholars note the effects vary a lot by context and task type, and causal chains from acute test anxiety to career abandonment are complex. Still, stereotype threat offers a plausible causal pathway for the negative outcomes examined in Research Question2.

Finally, this study uses the Stereotype Inoculation Model, proposed by Nilanjana Dasgupta (2011), to guide the evaluation of empowerment strategies to answer Research Question 3. According to this model, interacting with successful and relatable members of one's own peer such as female mentors or role models act as a psychological buffer against stereotypes. This helps protect an individual's self-concept from the negative effects of harmful stereotypes. The model directly supports RQ3, which examines how effective the empowering methods like mentorship and contact with role models work.

These broad theories align well with findings from German research, but there are local findings that require specific attention. Studies on German schools, for instance, show that teachers' and parents' gendered beliefs about MINT (Mathematik, Informatik, Naturwissenschaften und Technik; the German equivalent of STEM) subjects reduce girls' perceptions of those fields' usefulness and their intention to pursue them (Gaspard et al., 2015). Interviews and qualitative work with women in German computer science programs reveal recurring themes, these women often describe underestimating their abilities and feeling like outsiders in their field, that highlights stereotype threat (Breidenbach et al., 2021). In response, interventions within Germany have applied the Stereotype Inoculation Model in practical ways. One example is the large-scale e-mentoring program CyberMentor, which connects girls with female mentors. Research shows that this access increases girls' motivation and confidence in STEM, helping to protect them from the negative effects of stereotypes (Stoeger et al., 2013).

**Figure 1. A Conceptual Model of the Vicious Cycle of Gender Exclusion and Empowerment Interventions.**

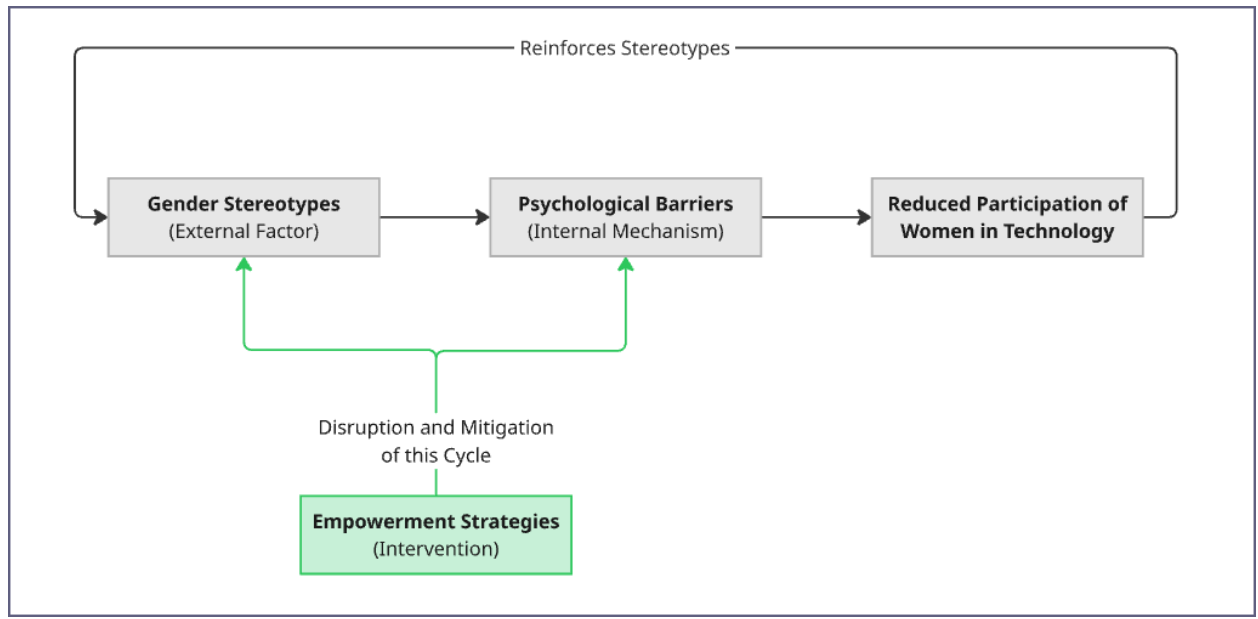

*Source: Authors' own elaboration, based on the synthesized literature.*

As illustrated in Figure 1, the present theoretical framework demonstrates that gender stereotypes, by generating psychological barriers, reduce women's participation in the technology sector. Conversely, empowerment strategies can disrupt this cycle and promote more sustainable female engagement in the field.

## Methodology

This study employs a structured literature review methodology (Petticrew & Roberts, 2006), this was selected as the most appropriate method to systematically identify, evaluate, and synthesize the evidence on the impact of gender stereotypes and psychological barriers of those stereotypes and the effectiveness of empowerment strategies for women in technology. The review was conducted and reported according to the Preferred Reporting Items for Systematic Reviews and Meta-Analyses (PRISMA) guidelines (Page et al., 2021). A structured review provides a transparent and strong way of processing and selection of literature; at the same time, it allows a flexible and thematic synthesis of findings to conduct a coherent argument at the end.

**Eligibility criteria**

Studies were selected for inclusion based on a pre-defined set of criteria structured around the Population, Exposure/Intervention, Outcome, and Study Design (PEOS) framework (Petticrew & Roberts, 2006).

- **Population**: Studies had to involve female participants (this included school-age girls, undergraduate students, and women working in technology, computer science, engineering, or adjacent STEM fields).

- **Exposure / Intervention**: Papers needed to report on at least one of the following: (a) psychological effects of gender stereotypes, (b) a psychological barrier commonly discussed in the literature (for example, stereotype threat, impostor phenomenon, or low task-specific self-efficacy), or (c) an empowerment-oriented intervention (examples in the literature range from sustained mentoring programs to short role-model talks or structured confidence-building workshops).

- **Outcomes:** The study must have reported outcomes relevant to the research questions. This included psychological precursors to participation, such as career interest, self-efficacy, and sense of belonging; key career-level outcomes like performance, persistence, and career aspirations; and measures of improvement in these areas following an empowerment intervention.

- **Study design and timeframe:** We did not limit by method: empirical quantitative and qualitative studies, as well as substantial review articles, were included. Searches were restricted to English-language publications between 2000 and 2025.

**Information sources and search strategy**

The search combined a broad database sweep with targeted follow-ups to capture influential studies.

- **Primary database search**: We ran structured queries in four major sources selected for complementary coverage: EBSCO Academic Search Ultimate (broad social science coverage), PsycINFO (psychology), ACM Digital Library and IEEE Xplore (computer science and engineering literature).

- **Search string**: A standardized Boolean string combined three concept groups. For example: (woman* OR female* OR girl*) AND (tech* OR "computer science" OR engineering OR STEM) AND (stereotype* OR "impostor" OR "impostor phenomenon" OR self-efficacy OR belonging OR mentor* OR "role model"). The actual queries were adapted to each database's syntax.

- **Supplementary searches**: To pick up highly cited or foundational works not prioritized by database algorithms, we cross-checked results in Semantic Scholar and used backward citation chasing (snowballing) from the reference lists of included papers (Wohlin, 2014).

**Study selection process**

Two authors carried out screening in stages to keep the process reproducible and to reduce individual bias. Each author first screened titles and abstracts independently; we then compared decisions and resolved disagreements through discussion and by re-consulting the inclusion criteria. Full texts were retrieved for items that passed the initial screen and assessed in detail.

The flow of records was as follows: an initial search returned 800 records. After removing 450 duplicates in a reference manager, 350 records remained and were screened by title/abstract, leading to 200 exclusions at that stage. We obtained the full texts of the remaining 100 articles; 42 of these were excluded for reasons such as lacking empirical evidence, or representing the wrong document type. The final narrative synthesis drew on 58 included articles.

**Data extraction and synthesis**

A thematic approach was used to synthesize the evidence from the final set of included articles (Braun & Clarke, 2006). We used a structured extraction form a "Synthesis Matrix" adapted from Garrard (2017) to capture the essentials from each study: research questions, sample characteristics, measures and instruments (when reported), key findings, theoretical framing, and any illustrative examples or participant quotes. Where papers reported intervention details, we recorded duration and format where that information was available.

For synthesis we adopted an iterative thematic approach to identify recurring patterns and dominant themes across the literature. These themes were organized according to the three primary research questions. The findings were then woven into the narrative synthesis presented in the "Findings" section, which builds a logical argument from the definition of the problem (stereotypes and barriers) to the evaluation of potential solutions (empowerment strategies).

# Findings

**The impact of gender stereotypes on women's career interest, self-efficacy**

Across the literature there is wide agreement that gendered ideas about technology help explain why women are underrepresented in computing and engineering. These ideas shape what children and young adults expect of themselves and of the field long before anyone makes a formal "career choice." Still, the story is not monolithic, there are cultural differences, individual exceptions, and places where interventions have nudged things in the other direction.

**The early and pervasive onset of stereotypes**

Stereotypes linking technical work to men show up very early. From toy aisles and birthday presents (think branded STEM kits, action-figure packaging, or construction sets aimed at boys) to the kinds of examples teachers pick in class and the characters children see on TV, everyday cues convey subtle messages: "tech is for boys." Several reviews and empirical studies document that these messages translate into beliefs, children as young as six sometimes already expect boys to be better at computer science and engineering (David, 2023; Olsson & Martiny, 2018; Wang & Degol, 2017; Miller et al., 2024). Importantly, some work suggests these technology-specific stereotypes appear earlier and stronger than comparable math stereotypes. Longitudinal evidence also reveal a significant challenge: rather than fading in adolescence, pro-male tech stereotypes can harden, with some cohorts of girls coming to endorse them more as they age (Miller et al., 2024). That said, trends vary by country and context, one Spanish study even reported an uptick in the perception of informatics and engineering as "male" over the last decade, a reminder that progress is neither steady nor uniform (Díez et al., 2022).

**The nature of tech stereotypes: a "masculine default"**

What keeps women out is not only questions about aptitude but also about fit. A recurring theme is the idea of a "masculine default" in tech cultures, such as environments, norms, and expectations that implicitly reward stereotypically masculine traits (Schmader, 2022). Physical surroundings matter, classic experiments found that a room decked out with clearly "geeky" paraphernalia (for example sci-fi posters, video game boxes, and collectible figurines) made women report less interest in computer science than when the room contained more neutral decor such as plants or general artwork (Prives, 2013). Media portrayals amplify this: the "asocial, lone-genius" programmer trope (often a white male character) is still common and can make the field seem emotionally uninviting to people who value collaborative, socially oriented work (Corsbie-Massay & Wheatly, 2022; Master & Meltzoff, 2020). Workplace expectations play a part too: ideals about the "perfect engineer" (available around the clock, willing to travel, unencumbered by family responsibilities) sit uneasily next to stereotypes that mothers are less committed, creating structural and perceptual barriers (Yates & Skinner, 2021).

**Psychological consequences: belonging, confidence, and interest**

Those cues add up. When an environment signals "you don't belong here," motivation drops, belonging is consistently identified as a key psychological gateway. A weak sense of fit often precedes disengagement. Confidence is also affected: many studies show that girls and women report lower STEM self-confidence than male peers even when test scores or grades are comparable (Martínez et al., 2023; Perera et al., 2024). Lower belonging and lower confidence then feed into career choices, so people tend to drift toward roles that feel aligned with their values and identities, and when tech is framed as individualistic or object-focused, it can feel incongruent with communal or socially oriented motivations that many women report (Wang & Degol, 2017).

Overall, then: gendered stereotypes create a network of small but persistent barriers, such as cultural signals, everyday interactions, and institutional expectations, that together make technology feel less welcoming to many women.

**The Impact of Psychological Barriers on Women's Persistence and Career Aspirations**

A large body of research shows that widely held gender stereotypes often get absorbed into women's everyday thinking and behavior, producing psychological barriers that can blunt performance, reduce persistence, and narrow career ambitions in technology. These are not just private worries; they tend to emerge predictably from navigating workplaces and classrooms where men are the visible norm. Three interlinked barriers appear most often in the literature: a weakened sense of belonging, falling confidence (often framed as the impostor phenomenon), and the tangible effects of stereotype threat.

**A diminished sense of belonging as a central barrier**

Across many studies a theme repeats: the environment signals who "belongs," and when those signals favor a masculine prototype, women notice. The cues can be small, for instance a conference room plastered with photos of male founders, team rituals that revolve around late-night "debugging beers," or social networks that form around engineering hobbies stereotypically coded male, yet they impact. Qualitative interviews vividly capture the experience: women describing what it feels like to be the only woman in a lab session, to hesitate before speaking in a sprint planning meeting, or to drift out of informal Slack threads where technical reputations are built (Master & Meltzoff, 2020; Prives, 2013; Phillips, 2024; Thackeray, 2016). This lack of "social fit" is identified as a key component of the "SAFE model," which argues that this mismatch between a woman's self-concept and the prevailing culture is a primary reason for attrition (Schmader, 2022).

**Eroded self-efficacy and the impostor phenomenon**

Confidence gaps show up repeatedly, even in cases where objective performance is comparable. Many papers document that female students or junior engineers report lower self-assessments in technical tasks than male peers despite similar grades or code reviews (Martínez et al., 2023; Brage-del-Río et al., 2025). People describe a relentless need to repeatedly demonstrate their capabilities: having to reassert competence in meetings, technical reviews, or when applying for roles. That cycle (external questioning feeding internal doubt) is closely tied to what scholars call the impostor phenomenon (Yates & Skinner, 2021; Kovaleva et al., 2022)

**Stereotype threat and its impact on performance**

Stereotype threat, the anxiety that comes from being aware of a negative stereotype about one's group, has clear experimental effects. For instance, studies where gender stereotypes are made salient before a math or coding task show short-term rises in anxiety and corresponding drops in measured performance (Yağan & Avcı, 2023). Practically speaking, that can mean a student who knows the stereotype walks into a timed exam with extra cognitive load, or an engineer who second-guesses choices during a live coding interview. Over time, repeated exposures to "chilly" or suspicious environments do more than depress single-test scores: they shape decisions about whether to apply for promotions, take on high-visibility projects, or remain in the field at all (Corsbie-Massay & Wheatly, 2022; David, 2023). Together, feeling like an outsider, doubting oneself, and worrying about confirming a stereotype tend to reinforce one another. The combined effect can function like a psychological disadvantage that makes technical careers harder to sustain for many women, reduces women's performance, lowers their persistence, and ultimately curtails their career aspirations in technology.

**The Effectiveness of Empowerment Strategies in Mitigating Barriers and Improving Outcomes**

The research agrees on one basic point: stereotypes and psychological barriers make it harder for many women to enter and stay in technology. At the same time, studies point to a handful of practical strategies that can help. How well they work depends on who runs them, where they are used, and whether they are sustained over time. Short, well-meaning interventions can help in the short term but often need follow-up to create lasting change.

**Role models and mentorship**

Seeing women who do well in tech matters. When girls and young women meet relatable role models, for example, a software engineer from a nearby startup who also juggled a young family, or a university lecturer who started coding as an adult learner, this can make a tech career feel attainable. Papers report fewer stereotype beliefs and higher confidence after such exposure (McCullough, 2011; Olsson & Martiny, 2018). But very distant "superstar" figures can feel out of reach and even discourage some students, whereas "ordinary" and accessible mentors are more impactful. Formal mentorship programs, especially those that give time for honest conversation (monthly one-to-one meetings, shadowing days at a company), tend to work more effectively. Formal mentorship programs, especially those within a single-gender environment, are shown to be highly effective at deconstructing stereotypes and reinforcing self-confidence by providing a benevolent and empathetic space for women to express their doubts (Bueno Merino & Duchemin, 2022; Shafiq et al., 2024).

**Peer support networks and belonging**

A second major theme is the critical importance of peer support networks for fostering a sense of belonging and building resilience. Qualitative work with female undergraduates describes small groups, such as study pairs, weekly coding meetups, or student chapters like a Women in Computing club, as key to pushing through rough patches (Phillips, 2024; Thackeray, 2016). Program evaluations back this up: buddy systems, peer tutoring, and community projects raise persistence and interest (Shafiq et al., 2024; Vaka et al., 2024). These findings suggest that creating "microenvironments" with a higher proportion of women is a powerful strategy to buffer against the negative psychological effects of being a minority in the field.

**Inclusive educational practices**

The literature points to set of specific educational strategies that can empower girls from early ages. One of the effective interventions teaching a growth mindset, which is the belief that intelligence and ability are not fixed and can be developed through putting effort. An experimental study found that a single "one-shot" growth mindset intervention was sufficient to eliminate the male-bias gender stereotype in children (Law et al., 2021), this finding was supported by other reviews that they recommend emphasizing effort over innate talent (Wang & Degol, 2017). Furthermore, the design of the curriculum itself is an impactful lever for empowerment. The evidence suggests that programs should be designed to be gender-neutral rather than specifically "feminized," that focus on the communal and socially beneficial applications of technology (e.g., sustainability). These are shown to be more motivating for women (Kovaleva et al., 2022). This includes creating inclusive classroom environments by removing "geeky" stereotypical cues and using gender-neutral language and imagery (Wang & Degol, 2017). Together, with combining providing role models, fostering peer networks, and designing inclusive educational experiences strategies, the literature suggests a clear and evidence-based pathway to dismantling psychological barriers and empowering more women to persist and succeed in technology.

**The Impact of Gender Stereotypes on Women's Career Interest, Self-Efficacy in Germany**

The synthesized literature provides a strong and consistent picture that gender stereotypes are a primary and persistent factor that cause the underrepresentation of women in German technology and STEM fields. These stereotypes are not abstract beliefs, they are actively manifested in the culture of academia and the workplace, so they shape women's self-perceptions and career aspirations from a young age.

**The "male-dominated" STEM culture and how it works**

A recurring finding in German research is the existence of a male-dominated STEM culture (Best et al., 2013). This means technical fields are commonly framed as masculine, and that framing shapes everyday behavior. For example, textbooks and school projects may highlight mechanical examples (cars, engines, robots), while classroom praise and informal mentoring often go to boys. Large cross-national work shows explicit and implicit links between "science/math" and men in Germany, stronger than those linking these fields to women (Froehlich et al., 2022). On campus and at work this plays out in quieter, cumulative ways: men interrupt women more often in meetings, technical credit is assigned unequally, and hiring panels may favor stereotypical traits. Qualitative surveys of female scientists report a familiar refrain that stereotypes about roles, double burdens of care and teaching, and structural power imbalances still shape career trajectories. (Kube et al., 2024).

**Early socialization and shaping of interest**

The evidence suggests many differences in interest have roots in childhood and adolescence through what some authors call "technical socialization." Goreth and Vollmer (2022) argue that early experiences explain much of the later gap in STEM interest. Practical examples include parents who buy electronics kits more often for boys, teachers who recommend advanced physics to male students, or careers guidance that presents engineering as a good fit for "practical" boys. Still, these patterns are not fixed. School programs that introduce girls to hands-on coding or female engineer role models sometimes reduce the gap, which suggests socialization can be changed rather than only blamed.

**Eroding Self-Efficacy and a Sense of Belonging**

A predictable psychological consequence of being in a stereotyped environment is lower self-efficacy and a weakened sense of belonging. German studies repeatedly find that girls and women rate their STEM abilities lower than objective measures often warrant (Hess et al., 2023). In practical terms, this looks like a woman at a technical seminar being the only one asked to take minutes or being repeatedly mistaken for administrative staff rather than for a researcher. A large-scale survey of female students at top German technical universities found that even when the women were socially integrated, a significant portion them felt professionally isolated, they reported a constant need to "continually have to prove own abilities" and felt "labelled with 'special status'" (Best et al., 2013). This constant questioning of their competence and abilities from external sources and from within, directly undermines self-efficacy. Furthermore, the experience of a male-dominated culture creates a diminished sense of belonging, which is a powerful psychological barrier. As Kube et al. (2024) found, women scientists in Germany consistently report being "taken less seriously" and having their ideas ignored, which reinforced the feeling that they are outsiders in their own field.

# Discussion

This study showed that the gender gap in the technology sector is the result of a combination of gender stereotypes, psychological barriers, and structural obstacles. This issue is not merely an individual or educational challenge, but a serious social and economic crisis. Persistent inequality in women's participation in technology not only hinders their individual development but also limits the capacity for innovation and development across society. These results underscore the need for a multidimensional and urgent perspective on this crisis. Although previous research has mainly focused on the formation of gender stereotypes during primary school (Miller et al., 2024; Olsson & Martiny, 2018), the role of family and the preschool environment in creating early attitudes and fears toward technology has been less examined. This research gap highlights the importance of addressing early experiences and family interactions.

**Comparison with previous studies**

The findings of this review are consistent with many prior studies. For example, Miller et al. (2024) showed that gender stereotypes in the technology domain form at very early ages (around six years old), a result that aligns with the work of Olsson and Martiny. The present review found that these stereotypes not only reduce girls' interest in technology but also substantially undermine their confidence and sense of belonging. These findings are also consistent with, and extend beyond, the results of Wang and Degol (2017) and Martiny, whose main focus was mathematics: here it became apparent that technology-related stereotypes are broader and more persistent.

The results of the present review can be better understood by drawing on several key theoretical frameworks. First, Expectancy-Value Theory (EVT) explains why gender stereotypes reduce girls' and women's interest in technology. According to this theory, individuals' decisions to enter a field depend on two factors: expectation of success and the value they attribute to that domain. Stereotypes that present technology as a "male" domain undermine both factors and inhibit the formation of sustained motivation.

Second, the concept of Stereotype Threat shows that mere awareness of these stereotypes can generate performance anxiety and affect the quality of women's learning and work. This issue is particularly pronounced in educational and workplace situations where men predominate.

Third, the impostor phenomenon explains why even successful and capable women often feel inadequate. This phenomenon not only undermines confidence but also increases the likelihood of leaving technological career paths. Thus, the theoretical interpretation of the results indicates that the gender gap in technology cannot be attributed solely to external factors; the internalization of stereotypes and psychological experiences also play a fundamental role.

**Implications**

Although numerous studies in recent years have addressed the gender gap in technology, the findings of the present review indicate that we still face a serious crisis. This crisis harms not only women's future careers but also societal capacity for innovation and development. Therefore, comprehensive, multi-level programs are needed that simultaneously address the educational, psychological, and structural dimensions of the problem.

**School and university**

Most existing research has focused on the role of schools and universities in forming or reinforcing gender stereotypes in technology. These findings underscore the importance of designing inclusive educational programs at these levels. However, attention must be paid to the fact that the role of family and the preschool environment in creating early attitudes and even anxieties about technology has been less studied. This gap suggests that educational interventions should begin at earlier ages, even before school entry.

From a practical perspective, measures such as introducing successful women as role models in classrooms, designing e-mentoring programs for girls, and reducing stereotype-related anxiety through formative (non-graded) assessments can significantly mitigate the effects of stereotypes (Master, 2021). A clear historical example of the absence of female role models in technology is Ada Lovelace, who in the nineteenth century wrote an algorithm for Charles Babbage's calculating machine and is therefore regarded as the first programmer in history. Neglecting such figures not only weakens the historical memory of women in technology but also limits inspiring opportunities for new generations of girls (Essinger, 2014).

**Workplace implications**

In the workplace and industry, one of the most important barriers to women's participation and retention is a male-dominated organizational culture and the lack of women in leadership positions. This inequality reduces women's sense of belonging and dampens motivation. Research shows that access to support networks and organizational mentoring programs can play an important role in boosting confidence and creating career opportunities (Hunt, 2015; Dasgupta, 2011). Hiring and promotion processes should be reviewed to prevent the reproduction of gender biases. In addition to ensuring cultural fit, the use of inclusive language in job advertisements and the creation of diverse and multi-voiced job roles can strengthen women's sense of belonging (Cheryan et al., 2017). Increasing the presence of women in managerial and decision-making roles not only helps change the workplace climate but also serves as an inspirational model for the next generation.

**Implications: Policy & National Programs (Germany)**

In Germany, women still hold a small share of technology jobs. Statistics indicate that only about 16–18% of professionals employed in ICT and MINT fields are women, which indicates a persistent gender gap.

To address this issue, several national policies and programs have been implemented, including:

- **Komm, mach MINT:** a national pact to attract girls to MINT subjects and to link education and industry.
- **Girls' Day:** a national day for girls to experience technical occupations directly.
- **CyberMentor:** the largest online mentoring program supporting girls in STEM over the long term.
- **Professorinnenprogramm:** a national program to increase the number of female professors at universities.
- **Fraunhofer TALENTA and the Ada-Lovelace Project:** support for female researchers' career paths and the presentation of successful role models.

Despite these initiatives, the shortage of women in technology remains clear and highlights the need for broader and more sustainable actions.

**Limitations**

This study has limitations that should be noted. First, the main focus was on quantitative articles, and qualitative studies were less represented in the present review. Second, a large portion of the data was obtained from Western countries, and therefore generalization of the results to other cultural contexts should be undertaken with caution. Also, due to methodological limitations, it was not possible to examine longitudinal trends and changes over time.

**Future research directions**

- **Preschool and family:** Most research has focused on schools and universities. However, it is necessary to investigate how gender stereotypes about technology form during sensitive developmental and psychological stages before school.
- **Cross-cultural comparative studies:** Most data come from Western countries such as Germany and the United States. It is essential to conduct research in different cultural contexts, such as developing countries, to understand how patterns may differ.
- **Emerging fields:** Most studies have concentrated on computer science and engineering. The role of women in data science, artificial intelligence, and other emerging technologies should also be examined.

- **Longitudinal research:** Future studies should conduct longitudinal research to track changes in women's attitudes and behaviors from childhood through entry into the labor market.
- **Intervention effectiveness:** Programs such as CyberMentor, MINT initiatives, and Girls' Day should be further evaluated in terms of real effectiveness (for example, the percentage of participants who later enter technology fields).
- **Creating educational programs or academic tracks:** To accelerate effectiveness in eliminating the underrepresentation of women, trained human resources should be established whose responsibility is to work on the three main areas: gender stereotypes, psychological fears, and organizational change to empower women in technology occupations.

## Conclusion

This study shows that the gender gap in technology is not an ordinary issue but a global crisis whose consequences extend beyond the technology sector itself. The underrepresentation of women in technological fields means a loss of innovative capacity that could play a key role in addressing fundamental global challenges, from sustainable development to environmental and health problems.

Although multiple interventions and programs have been implemented at the levels of schools, workplaces, and national policy, women's presence in technological fields remains significantly limited. Addressing this problem requires a sustainable, multi-level approach that includes not only early education, the introduction of female role models, inclusive organizational practices, and structural reforms, but also attention to psychological interventions. Such interventions can, by reducing stereotype threat, strengthening self-efficacy, and countering self-undervaluation, pave the path to women's empowerment.

Based on these findings, the authors intend to develop this line of research through an interdisciplinary initiative entitled the "**NEURON**" project, an initiative aimed at designing innovative strategies to empower women in the field of technology.